\documentclass[twocolumn]{aa}
\usepackage{graphicx}

\usepackage{txfonts}
\begin{document}
\title{Testing Newtonian gravity in the low acceleration regime
with globular clusters: the case of $\omega$ Centauri revisited. }

   \subtitle{}

   \author{Riccardo Scarpa\inst{1} and Renato Falomo \inst{2}}

\offprints{R. Scarpa; riccardo.scarpa@gtc.iac.es}

\institute{Instituto de astrof\'isica de canarias, 
c/via Lactea s/n, San Cristobal de la Laguna 38205, Spain
\and INAF, Osservatorio astronomico di Padova, 
vicolo Osservatorio 5, Padova, Italy }

\date{\today}
\authorrunning{Scarpa \& Falomo}
\titlerunning{Testing Newtonian gravity in the low acceleration regime:
the case of $\omega$ Centauri revisited.}

\abstract{Stellar kinematics in the external regions of globular
  clusters can be used to probe the validity of Newton's law in the
  low acceleration regimes without the complication of non-baryonic
  dark matter. Indeed, in contrast with what happens when studying
  galaxies, in globular clusters a
  systematic deviation of the velocity dispersion profile from the
  expected Keplerian falloff would provide indication of a breakdown
  of Newtonian dynamics rather than the existence of dark matter.}
{ We perform a detailed analysis of the velocity dispersion in
  the globular cluster $\omega$ Centauri in order to investigate whether 
  it does decrease monotonically with distance as recently 
  claimed by Sollima et al. (2009),
  or whether it converges
  toward a constant value  as claimed by Scarpa Marconi and Gilmozzi (2003B).}
{ We combine  measurements from these two works to almost
  double the data available  at large radii, in this way obtaining
  an improved determination of the velocity dispersion profile 
  in the low acceleration regime.}
{ We found the inner region of $\omega$ Centauri is clearly rotating,
  while the rotational velocity tend to vanish, and is consistent with
  no rotation at all, in the external regions.  The cluster velocity
  dispersion at large radii from the center is found to be sensibly
  constant.}
{ The main conclusion of this work is that strong similarities are
  emerging between globular clusters and elliptical galaxies, for in
  both classes of objects the velocity dispersion tends to remain
  constant at large radii. In the case of galaxies, this is ascribed
  to the presence of a massive halo of dark matter, something
  physically unlikely in the case of globular clusters. Such
  similarity, if confirmed, is best explained by a breakdown of
  Newtonian dynamics below a critical acceleration.   }

\maketitle

%

\section{Introduction}

A fundamental aspect of our knowledge of the Universe concerns the
existence of non-baryonic dark matter (DM), believed to represents
about 20\% of the total energy budget of the Universe.  Signatures of
DM are found in galaxies and cluster of galaxies due to
its dynamical effects and gravitational lensing.  While DM appears in
wildly variable quantities and distributions among different types of
objects, it appears to exhibit systematic (but not yet
understood) behaviours (c.f., recent findings by \cite{gentile09}
and by Donato et al. 2009).  The most remarkable (e.g.,
\cite{binney04}) being that DM is needed to reconcile the observations
with the expectations of Newtonian dynamics when and only when the
acceleration of gravity goes below a {\it critical} value, 
$a_0 \sim 1.2 \times 10^{-8}$ cm s$^{-2}$ (\cite{begeman91}).

\begin{figure*}
\centering
\includegraphics[height=9cm]{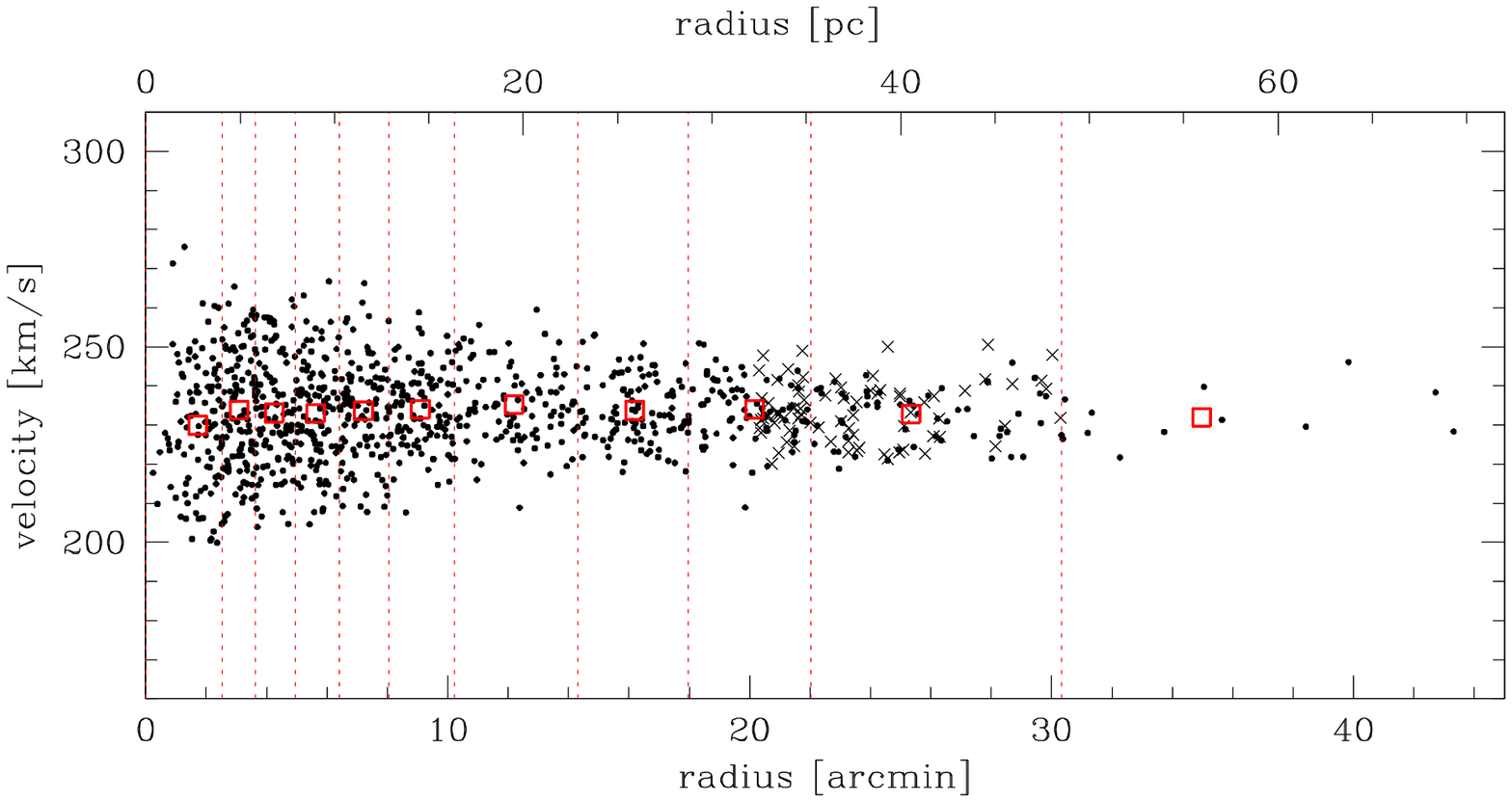}
\caption{\label{velocities} The distribution of radial velocities from
  \cite{sollima09} (points) as a function of distance from cluster
  center.  The sample of \cite{scarpa03B} (crosses), limited to the
  region $20<R<30$ arcmin, almost double the number of points in the 
  region where $a<a_0$.  The vertical (dotted) lines give the limits of bins
  containing 100 points each (beside the last one that contain only 13
  points).  The average radial velocity within each bin (large open squares) is
  very stable with no indication of peculiar trends. The uncertainty
  on the average is smaller than the size of the symbols.  }
\end{figure*}

This fact led to suggest that Newtonian dynamic might not be
applicable below this acceleration and the most successful proposal of
this type, known as MOND (\cite{milgrom83}), nicely explains the
rotation curves of spiral galaxies and many other dynamical properties
of galaxies without the needs for DM (\cite{mcgaugh98};
\cite{mortlock01}; see \cite{sanders02} for a review). Such an
hypothesis, however, has so many serious consequences for the standard
physics (e.g. \cite{milgrom09} for a review), that as many test as
possible should be carried out to verify it.  Experiments can be made
in the laboratory or studying astrophysical systems where DM is
absent.  A first pioneering study along this line focused on the
dynamics of the external regions of globular clusters, the largest
virialized structures that do not contain significant amount of dark
matter.  Measurement of the velocity dispersion in the outskirt of
$\omega$ Centauri (\cite{scarpa03A},B), showed a clear flattening of
the velocity dispersion profile starting at the radius where the
cluster's internal acceleration of gravity is $\sim a_0$, with no
evidence of the expected Keplerian falloff.  The very same behaviour
observed in elliptical galaxies and explained
invoking the presence of large amounts of dark matter.  This result was
then extended to other 6 globular clusters (Scarpa, Marconi, and
Gilmozzi 2003 A,B; Scarpa, Marconi, and Gilmozzi 2004 A,B; Scarpa et
al. 2007 A,B; Scarpa et al. 2010), showing the behaviour seen
in $\omega$ Cen is not a peculiar property of this cluster.

Given the relevance of this result, the dynamics of $\omega$ Cen was
recently carefully reconsidered by Sollima et al. (2009, S09
hereafter).  Based on the analysis of a new large dataset of radial
velocities measurements, it was claimed that the velocity dispersion
decreases monotonically with radius, in agreement with Newtonian
prediction, and in clear contrast with the claim by Scarpa Marconi and
Gilmozzi (2003B; SMG hereafter). We reconsider here the data
presented by S09. The reanalysis is done joining both the S09 and SMG
radial velocities data to create a larger sample.  In both works, the
selection criteria for cluster members identification was basically
the same (a selection based on position in the color
magnitude diagram combined with  a cut in  radial velocity), ensuring 
that the whole dataset is homogeneous.

\begin{figure}
\centering
\includegraphics[height=15cm]{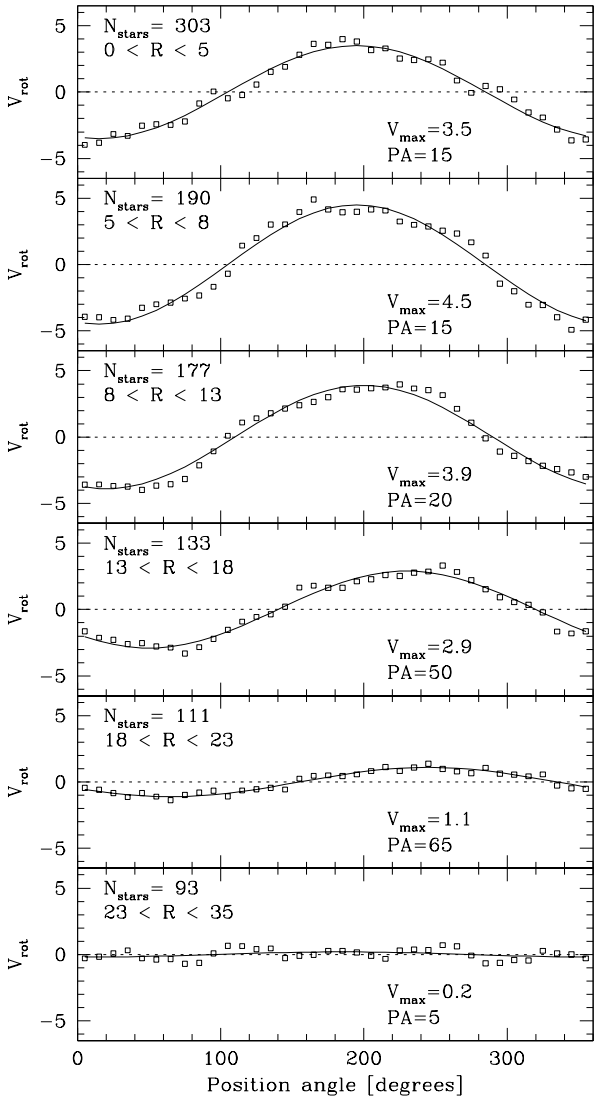}
\caption{\label{rotation} Rotation (in km/s) of
  $\omega$ Cen in 6 annular regions at increasing
  distance from the center.  The number of stars in each region and
  the radial distance limits (in arcmin)  are shown. 
  The formal value of the maximum 
  velocity, V$_{max}$, and the position angle, are derived from
  the best sinusoidal fit to the data (solid line). V$_{max}$ is given in
  km/s and the position angle in degrees from North toward East.}
\end{figure}

\begin{table}[b]
\label{tabCluster}
\scriptsize
\caption{$\omega$ Centauri (NGC 5139) basic properties}
\begin{tabular}{lll}
\hline
RA, DEC (2000) &  13:26:45.76,  -47:28:42.8  &Coordinates of cluster center\\
L,B            &    309.10  14.97  &Galactic coordinates\\
R$_{sun}$  &    5.5  kpc & Distance from sun\\
R$_{MW}$   &  6.4  kpc & Distance from Milky Way center\\   
M$_V$ & $-10.29$   &Total V band magnitude\\
Mass/M$\odot$  & 1.1$\times 10^6$ &From luminosity assuming M/L$_V$=1\\ 
r$_e$ &4.8 arcmin or  7.7 pc & Half light radius\\
r$_t$ &45 arcmin or 72 pc & Tidal radius\\
 Scale factor & 1.60 &pc/arcmin \\
\hline
\hline
\multicolumn{3}{l}{Coordinates are from van de Ven 2006. The exact cluster center used in S09}\\
\multicolumn{3}{l}{is 13:26:46.5, -47:28:41.1, Sollima private communication. This minimal }\\
\multicolumn{3}{l}{difference has no effects on our analysis.}\\
\end{tabular}
\end{table}

\section{Radial velocity measurements for $\omega$ Centauri}

Located at 6.4 kpc from the galactic center (\cite{harris96}),
$\omega$ Cen  is the most massive and luminous globular cluster of the
Milky Way.  It is sufficiently massive to contain more than one
stellar population, as indicated by helium abundance variation (Norris
2004; Piotto et al. 2005) and its peculiar position in the
size-luminosity plane (\cite{mackey05}).  It has been also argued that
$\omega$ Cen could be not a genuine globular cluster but the nuclear
remnant of a dwarf galaxy that merged in the past with the Milky Way
(e.g., \cite{bellazzini08}).

Among the earliest dynamical studies of $\omega$ Cen relevant to this
work, Meylan \& Mayor (1986) discussed the radial velocities of 318
cluster members, covering the cluster up to 22 arcmin from the
center. Comparing these data to the velocity dispersion derived from
proper motion data for several thousands stars (\cite{vanLeeuwen00}),
it was possible to demonstrate (\cite{scarpa03B}) that the velocity
ellipsoid is isotropic. A very important result for our purposes,
because at large radii only the velocity dispersion along the line of
sight can be measured.

Assuming for the cluster a total absolute magnitude of M$_V=-10.29$,
distance of 5.5 kpc \footnote{For consistency we use here a distance
of 5.5 kpc as in S09, while in SMG an older
value of 5.1 kpc (van Leeuwen et al. 2000) was used.}, and
mass-to-light ratio M/L=1 in solar units, the acceleration is
$a_0$ at $r_0 \sim 22.3$ arcmin. Thus to extend the results presented
by Meylan \& Mayor (1986) to radii where the acceleration goes below
$a_0$, SMG obtained radial velocity measurements for 75
cluster members in the region $20<r<30$ arcmin.
The full list of radial velocities obtained by SMG was
not published at that time, thus it is given in Table 2. Details on
the observations and analysis of individual measurements are reported
in SMG. The work by SMG, while confirming the results
by Meylan and Mayor in the region of overlap, showed that the velocity
dispersion did not decrease with distance beyond $r \sim 20$ arcmin.

A comprehensive study of this cluster was then presented in van de Ven
2006, discussing both proper motions and radial velocities.  The data,
however, did cover the cluster only up to 20 arcmin from the
center. The velocity dispersion profile was found fully consistent
with earlier determinations.

In S09 radial velocities measurements for 946 cluster members, of
which 628 originally presented by \cite{Pancino07}, probing the
cluster up to $r\sim 45$ arcmin from the center were discussed.  The
sample included 98 data points in the $20<r<45$ arcmin region
(Fig. \ref{velocities}), comparable to the number of measurements of
SMG.  Sollima and collaborators found the velocity dispersion to be
decreasing up to $\sim 26$ arcmin from the center reaching a minimum
of $5.2$ km/s.  Outward, at 32 arcmin, they measured a dispersion of
$7$ km/s. They regarded this increase ``more compatible with the onset
of tidal heating that with the effects of MOND or DM'', and went on
with the conclusion that the velocity dispersion in $\omega$ Cen is
fully compatible with Newtonian dynamics.

\section{The rotation of $\omega$ Centauri}

$\omega$ Cen is one of the most flattened globular clusters known, suggesting
a significant rotation component. 
Before discussing the velocity dispersion, it is therefore important to quantify
the fraction of the total energy budget of $\omega$ Cen that goes into
ordered motion.  Clear evidence of rotation in the inner regions of the cluster
was found by Meylan and Mayor (1986). More recently Pancino et al. (2007)
using  a fraction of the S09 sample, measured a maximum rotation of 6.8 km/s
between 6 to8 arcmin from the center. Using the large SMG + S09 combined
samples, we evaluated the amount of rotation in 6 regions at
increasing distance from the center.  In each region we halved the cluster by
position angle and compute the mean radial velocity of each half.  The
difference of these two velocities correspond to twice the rotational
velocity.  This procedure was applied with steps of 10 degrees in
position angle (Fig. \ref{rotation}).

The center of the cluster is clearly rotating. In the two innermost
bins the formal value of the maximum rotation velocity is 4.5 km/s,
somewhat lower than quoted by Pancino et al. (2007), with rotation
axis position angle of 15 degrees (from North toward East). The
rotational velocity starts decreasing at $r \sim 10$ arcmin, also
showing indication of a possible drift of the rotation axis.  In the
two most external annulus, the rotational velocity is very small,
being consistent with no rotation at all in the outer most bin.

In all six regions, the rotational velocity is
significantly smaller than the velocity dispersion (see next section),
indicating that the amount of energy stored into ordered rotational
motion is negligible compared to the one of the chaotic motion. Thus,
at large radii the velocity dispersion should closely follow a typical
Keplerian falloff (unless external effects modify it).

\begin{figure}
\centering
\includegraphics[height=11.5cm]{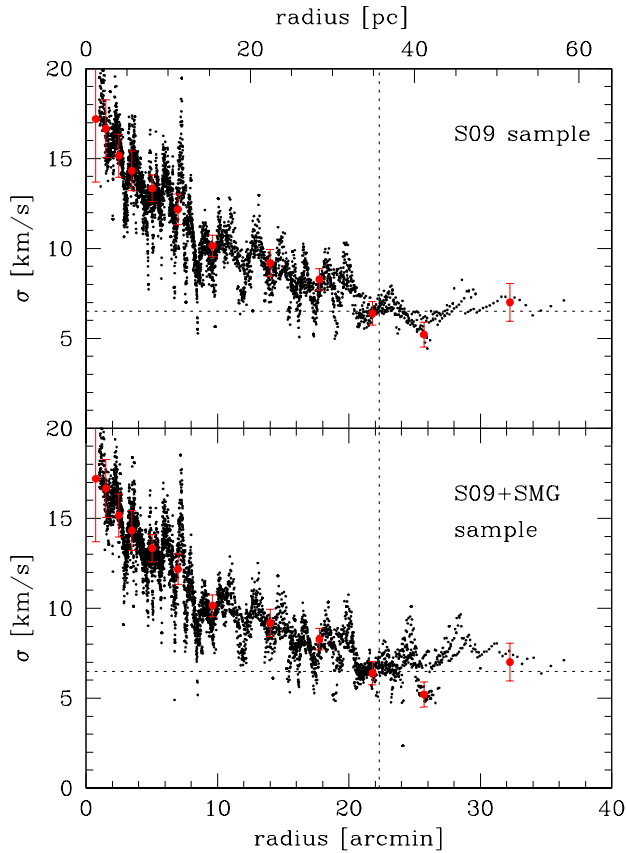}
\caption{\label{dispFixBin} Plot of 125 velocity dispersion profiles,
  each one obtained with a different binning of the data from  the
  S09 sample alone (upper panel), or the combined S09+SMG sample
  (lower panel).  The envelope of all these profiles, rather than the
  singular points,  shows the flattening of the profile beyond
  $r\sim 20$ arcmin.  The velocity dispersion reported by S09 (red
  points with error bars) nicely fall close to the center of the
  envelope. However, the value of $\sigma=5.2$ km/s at $r=25.7$ arcmin is
  not representative of the data, being in both cases at the very
  bottom of the envelope. The vertical line gives the MOND radius,
  while the horizontal one is only meant to highlight the flattening
  of the dispersion profile.}
\end{figure}

\begin{figure}
\centering
\includegraphics[height=10.5cm]{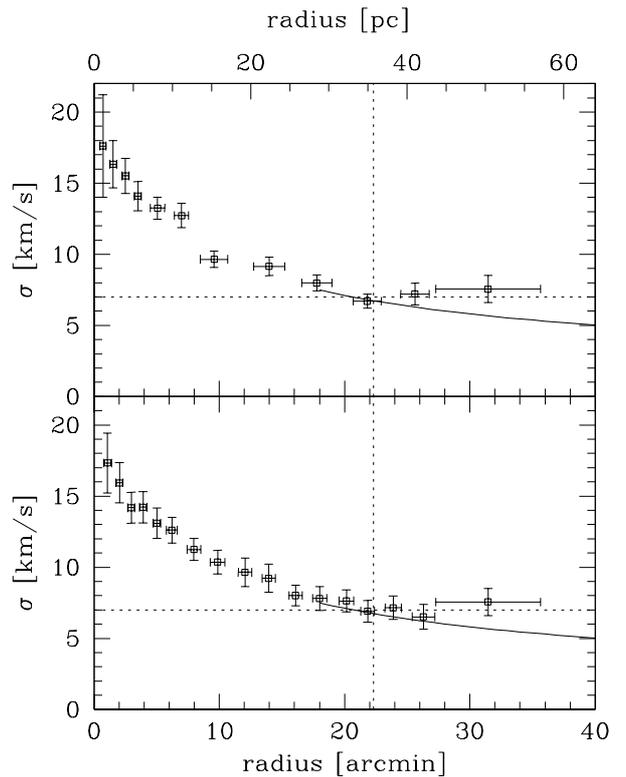}
\caption{\label{veldisp} The velocity dispersion profile of $\omega$
  Cen as obtained considering the combined S09+SMG sample, and two
  different binning.  {\bf Top panel:} Data binned as in S09.  
  {\bf Bottom panel:} Narrower sampling for having one extra bin beyond
  the MOND radius.  Points give the velocity dispersion in km/s and
  its 1$\sigma$ uncertainty, while the bars in the x direction give
  the rms of the data in the bin. The vertical dotted line indicates
  the MOND radius, where the acceleration is $a_0$. The horizontal
  line is not a fit to the data, it is only meant to highlight the
  flattening of the dispersion profile. The solid line show a 
  Keplerian fall off. Note that all these data points fall well within 
  the cluster tidal radius (45 armnin). 
}
\end{figure}

\section{The velocity dispersion profile of $\omega$ Cen.}

It is well known that binning of sparse and non homogeneously
distributed data can produce artificial/spurious trends in binned
plots.  A bin independent method of constraining the velocity
dispersion is shown in Fig. \ref{dispFixBin}, where we plot on top of
each other 125 velocity dispersion profiles, each one obtained 
with a different binning.

To create these profiles, we first sorted the data according to
increasing distance from the cluster center. Then we constructed
binned data starting from point number 10, 20, 30, 40, and 50,
disregarding the innermost points that are irrelevant for our
purposes. The number of points per bin, constant for each profile, was
varied from a minimum of 10 to a maximum of 50 points, in step of
2. This procedure resulted in 125 different combinations of binned
profiles.  What is relevant for the description of the velocity
dispersion profile is the envelope of all the profiles.  While the
dispersion varies significantly from point to point, the flattening at
large radii is evident, both when using data from the S09 sample
alone, and the S09+SMG combined samples. The values computed by S09
(see their Table 1) are very close to the center of the envelope, but
the point at $r=25.7$ ($\sigma=5.2$) is clearly not representative of
the data, being at the bottom of the envelope.  It seems to us S09
made an unfortunate choice of the binning, that resulted in an
underestimation of the velocity dispersion in the critical region
beyond 22 arcmin.  It turns out therefore, that both the S09 and the
combined S09+SMG data exhibit a clear flattening (at $\sigma\sim 7$
km/s) of the velocity dispersion profile in $\omega$ Cen at large
radii ($r>20$ arcmin).

We present in Fig. \ref{veldisp} the velocity dispersion obtained
binning the combined S09+SMG sample with the S09 original binning, and
with a narrower binning that better samples the profile (see Tab. 3).
The expected Keplerian fall off is also shown. This was computed
assuming all the cluster mass is contained within the MOND radius, an
assumption justified by the fact that 95\% of the cluster total
luminosity is contained within a radius as small as 17 arcmin
(\cite{trager95}).  Although the departure of the velocity dispersion
profile from the Keplerian fall off is modest it shows a systematic
trend.

\section{Discussion}

The analysis of the largest available data set of radial velocities of
stars in the globular cluster $\omega$ Cen, show the velocity
dispersion at large radii from the center remains constant
(Fig. \ref{veldisp}). There is indeed no indication of the decrease of
the velocity dispersion claimed by S09.  The flattening starts at $r=
20\pm 2$ arcmin, where the Newtonian acceleration of gravity is $a =
\frac{GM}{r^2} \sim 1.5\times 10^{-8}$ cm s$^{-2}$ in line with the
values measured in the other globular clusters studied as part of this
project (Scarpa et al. 2007A, Scarpa et al. 2010). Here G is the gravitational
constant and the mass M 
is computed assuming a mass-to-light ratio M/L=1 in solar
units, and a cluster total absolute magnitude of M$_V=-10.29$
(\cite{harris96}), further assuming all the mass is contained
  within the MOND radius, an approximation well within the
  uncertainties of the M/L ratio.

Although, due to the still limited statistics a Keplerian falloff at
large radii cannot be ruled out in $\omega$ Cen, combining
the results for all globular clusters studied so far there is mounting
evidence that globular clusters mimic nicely the behaviour of
elliptical galaxies (Scarpa et al.  2010).

Whether this implies this is the effect of MOND in a class of
objects that do not contain dark matter remains unclear.  Indeed, it
was argued (\cite{baumgardt05}) that $\omega$ Cen is too close to
the Milky Way to provide a useful test for MOND.  Milgrom (1983)
explicitly stated that  due to the strong external field of the Milky
Way no deviations from a Keplerian fall off should be observed in
$\omega$ Cen (this is the case for all the other clusters we studied
so far). 

Alternatively one might assume M/L$>1$ so that the acceleration
remains above the MOND threshold at all radii probed by the data. 
For instance Moffat and Toth
(2008) using M/L=2.9 obtained in the framework of Newtonian dynamics
a reasonable fit of the dispersion
profile from SMG. It would be
interesting to see whether it is still possible to fit the current
data that exhibits an even more marked flattening.  We note, however, 
that the
assumptions of M/L significantly larger than 1 is rather
controversial, because in the few cases where the present-day mass
function of globular clusters has been measured, M/L is 
found to be even smaller than 1
(\cite{demarchi99}; \cite{piotto97}; \cite{andreuzzi00}).

Irrespectively of which is the correct theoretical interpretation of
the flattening of the velocity dispersion profile, we believe a
striking similarity between globular clusters and elliptical galaxies
is emerging.

We thanks Sollima and collaborators for giving us access to
their full data sample. This work was supported by ASI-COFIC 
contract n. I/016/07/0 "Studi di Cosmologia e Fisica Fondamentale".


\begin{table}
\label{velSMG}
\scriptsize
\caption{Heliocentric Radial velocities for $\omega$ Centauri members }
\begin{tabular}{cccc}
 ID &   RA   &  DEC     & Velocity   \\
    & (2000) & (2000)   & [km/s]     \\
\hline
 00006 &$  13:25:06.00 $&$ -47:09:21  $&$  222.62 \pm  0.51 $ \\ 
 01008 &$  13:25:14.88 $&$ -47:09:51  $&$  222.55 \pm  0.87 $ \\ 
 01009 &$  13:25:49.67 $&$ -47:09:49  $&$  244.38 \pm  0.00 $ \\ 
 02005 &$  13:25:15.12 $&$ -47:10:18  $&$  242.54 \pm  0.12 $ \\ 
 05007 &$  13:25:37.91 $&$ -47:11:29  $&$  230.56 \pm  2.44 $ \\ 
 05008 &$  13:25:40.93 $&$ -47:11:40  $&$  228.03 \pm  1.05 $ \\ 
 06009 &$  13:25:35.46 $&$ -47:12:00  $&$  235.71 \pm  0.00 $ \\ 
 08003 &$  13:24:16.22 $&$ -47:13:11  $&$  239.34 \pm  1.07 $ \\ 
 08004 &$  13:24:17.07 $&$ -47:13:14  $&$  241.36 \pm  0.75 $ \\ 
 10006 &$  13:24:39.15 $&$ -47:14:00  $&$  237.46 \pm  1.27 $ \\ 
 10009 &$  13:24:58.69 $&$ -47:14:02  $&$  227.59 \pm  1.13 $ \\ 
 10010 &$  13:25:20.30 $&$ -47:14:00  $&$  220.13 \pm  1.27 $ \\ 
 13006 &$  13:24:32.09 $&$ -47:15:31  $&$  231.72 \pm  0.69 $ \\ 
 14002 &$  13:24:38.98 $&$ -47:15:52  $&$  223.74 \pm  0.54 $ \\ 
 15007 &$  13:25:06.38 $&$ -47:16:22  $&$  222.83 \pm  0.25 $ \\ 
 16003 &$  13:24:03.53 $&$ -47:16:48  $&$  247.97 \pm  0.70 $ \\ 
 19005 &$  13:24:12.87 $&$ -47:17:53  $&$  224.55 \pm  1.91 $ \\ 
 20006 &$  13:24:34.17 $&$ -47:18:30  $&$  250.01 \pm  0.00 $ \\ 
 22007 &$  13:24:20.45 $&$ -47:19:35  $&$  226.94 \pm  0.90 $ \\ 
 22008 &$  13:24:59.82 $&$ -47:19:16  $&$  229.30 \pm  0.00 $ \\ 
 24011 &$  13:24:50.82 $&$ -47:20:09  $&$  228.63 \pm  0.96 $ \\ 
 25004 &$  13:24:44.87 $&$ -47:20:45  $&$  230.58 \pm  0.52 $ \\ 
 26009 &$  13:24:47.16 $&$ -47:21:14  $&$  224.52 \pm  0.01 $ \\ 
 27008 &$  13:24:39.61 $&$ -47:21:47  $&$  237.51 \pm  1.24 $ \\ 
 28009 &$  13:24:32.94 $&$ -47:21:53  $&$  235.90 \pm  0.62 $ \\ 
 31006 &$  13:24:30.90 $&$ -47:23:35  $&$  232.55 \pm  0.00 $ \\ 
 33006 &$  13:24:30.77 $&$ -47:24:26  $&$  225.94 \pm  1.69 $ \\ 
 34008 &$  13:24:46.79 $&$ -47:24:48  $&$  232.19 \pm  0.81 $ \\ 
 37009 &$  13:24:38.53 $&$ -47:26:19  $&$  234.19 \pm  0.23 $ \\ 
 39013 &$  13:24:38.58 $&$ -47:27:03  $&$  231.71 \pm  0.90 $ \\ 
 42009 &$  13:24:26.66 $&$ -47:28:19  $&$  223.46 \pm  1.60 $ \\ 
 42012 &$  13:24:41.85 $&$ -47:28:29  $&$  232.86 \pm  1.11 $ \\ 
 43002 &$  13:24:34.11 $&$ -47:28:44  $&$  229.36 \pm  0.53 $ \\ 
 45011 &$  13:24:26.25 $&$ -47:29:38  $&$  224.22 \pm  0.58 $ \\ 
 45014 &$  13:24:37.49 $&$ -47:30:00  $&$  248.95 \pm  0.47 $ \\ 
 46003 &$  13:24:31.06 $&$ -47:30:09  $&$  241.79 \pm  0.51 $ \\ 
 48009 &$  13:24:28.89 $&$ -47:31:06  $&$  223.06 \pm  0.78 $ \\ 
 49008 &$  13:24:38.96 $&$ -47:31:37  $&$  238.58 \pm  0.83 $ \\ 
 51005 &$  13:24:22.05 $&$ -47:32:22  $&$  221.38 \pm  0.69 $ \\ 
 57006 &$  13:24:44.54 $&$ -47:35:06  $&$  234.32 \pm  0.38 $ \\ 
 61009 &$  13:24:38.47 $&$ -47:36:58  $&$  239.66 \pm  0.47 $ \\ 
 64010 &$  13:24:44.13 $&$ -47:38:21  $&$  225.80 \pm  2.67 $ \\ 
 72007 &$  13:25:02.53 $&$ -47:41:51  $&$  236.44 \pm  1.02 $ \\ 
 75005 &$  13:24:39.87 $&$ -47:43:21  $&$  235.89 \pm  3.84 $ \\ 
 76015 &$  13:25:21.21 $&$ -47:44:03  $&$  241.52 \pm  0.75 $ \\ 
 77010 &$  13:24:56.51 $&$ -47:44:28  $&$  238.13 \pm  0.78 $ \\ 
 78004 &$  13:24:14.39 $&$ -47:45:00  $&$  231.88 \pm  0.76 $ \\ 
 78008 &$  13:24:52.57 $&$ -47:44:48  $&$  222.95 \pm  0.57 $ \\ 
 79008 &$  13:25:03.49 $&$ -47:45:27  $&$  238.80 \pm  1.00 $ \\ 
 80017 &$  13:25:36.42 $&$ -47:45:31  $&$  231.63 \pm  0.75 $ \\ 
 80019 &$  13:25:39.47 $&$ -47:45:49  $&$  236.88 \pm  1.46 $ \\ 
 82012 &$  13:25:46.06 $&$ -47:46:45  $&$  232.43 \pm  0.57 $ \\ 
 85007 &$  13:24:46.16 $&$ -47:48:02  $&$  250.58 \pm  0.37 $ \\ 
 85014 &$  13:25:30.54 $&$ -47:48:04  $&$  236.89 \pm  0.82 $ \\ 
 85019 &$  13:26:09.52 $&$ -47:48:06  $&$  244.01 \pm  0.64 $ \\ 
 86007 &$  13:24:53.73 $&$ -47:48:16  $&$  238.73 \pm  0.80 $ \\ 
 86010 &$  13:25:14.92 $&$ -47:48:28  $&$  238.07 \pm  0.61 $ \\ 
 86017 &$  13:26:14.90 $&$ -47:48:27  $&$  231.25 \pm  1.41 $ \\ 
 87009 &$  13:26:28.10 $&$ -47:48:58  $&$  247.74 \pm  0.64 $ \\ 
 88022 &$  13:26:17.91 $&$ -47:49:27  $&$  225.56 \pm  0.00 $ \\ 
 88023 &$  13:26:20.45 $&$ -47:49:13  $&$  233.11 \pm  0.52 $ \\ 
 89009 &$  13:26:16.96 $&$ -47:49:58  $&$  242.16 \pm  0.53 $ \\ 
 89014 &$  13:26:39.70 $&$ -47:49:51  $&$  231.79 \pm  0.44 $ \\ 
 90008 &$  13:24:53.91 $&$ -47:50:24  $&$  240.48 \pm  0.44 $ \\ 
 90019 &$  13:26:30.06 $&$ -47:50:17  $&$  232.78 \pm  0.49 $ \\ 
 90020 &$  13:26:34.47 $&$ -47:50:17  $&$  240.16 \pm  0.62 $ \\ 
 91010 &$  13:25:20.97 $&$ -47:50:34  $&$  227.18 \pm  2.77 $ \\ 
 93016 &$  13:26:36.13 $&$ -47:51:43  $&$  230.91 \pm  0.67 $ \\ 
 94011 &$  13:25:14.21 $&$ -47:51:53  $&$  241.77 \pm  0.58 $ \\ 
 94014 &$  13:26:31.35 $&$ -47:51:55  $&$  229.29 \pm  0.10 $ \\ 
 95013 &$  13:25:58.71 $&$ -47:52:26  $&$  237.44 \pm  0.53 $ \\ 
 95015 &$  13:26:11.78 $&$ -47:52:22  $&$  238.97 \pm  0.53 $ \\ 
 96011 &$  13:26:05.55 $&$ -47:52:55  $&$  229.73 \pm  0.65 $ \\ 
 97012 &$  13:26:06.65 $&$ -47:53:12  $&$  233.22 \pm  0.00 $ \\ 
 98012 &$  13:25:25.27 $&$ -47:53:46  $&$  229.82 \pm  0.67 $ \\ 
\hline
\hline
\end{tabular}
\end{table}

\begin{table}[b]
\label{tabdisp}
\scriptsize
\caption{Radial velocity dispersion for $\omega$ Centaury}
\begin{tabular}{crrr}
Bin limits  &  R~~~~~~    & N   & Dispersion \\
  (arcmin)  & (arcmin) &     & km/s~~~~~  \\
\hline
    $       0    -     1.5  $ &     1.07   &       34   & $   17.33  \pm  2.11 $\\
    $     1.5    -     2.5  $ &     2.03   &       63   & $   15.94  \pm  1.42 $\\
    $     2.5    -     3.5  $ &     2.98   &       86   & $   14.18  \pm  1.08 $\\
    $     3.5    -     4.5  $ &     3.92   &       83   & $   14.22  \pm  1.11 $\\
    $     4.5    -     5.5  $ &     5.01   &       75   & $   13.09  \pm  1.07 $\\
    $     5.5    -       7  $ &     6.22   &      100   & $   12.61  \pm  0.89 $\\
    $       7    -       9  $ &     7.97   &      104   & $   11.25  \pm  0.78 $\\
    $       9    -      11  $ &     9.87   &       78   & $   10.35  \pm  0.83 $\\
    $      11    -      13  $ &    12.06   &       47   & $    9.64  \pm  1.00 $\\
    $      13    -      15  $ &    13.95   &       44   & $    9.21  \pm  0.99 $\\
    $      15    -      17  $ &    16.09   &       64   & $    8.01  \pm  0.71 $\\
    $      17    -      19  $ &    18.01   &       44   & $    7.80  \pm  0.84 $\\
    $      19    -      21  $ &    20.14   &       50   & $    7.63  \pm  0.77 $\\
    $      21    -      23  $ &    21.85   &       42   & $    6.90  \pm  0.76 $\\
    $      23    -      25  $ &    23.88   &       39   & $    7.14  \pm  0.82 $\\
    $      25    -      28  $ &    26.30   &       28   & $    6.51  \pm  0.88 $\\
    $      28    -      44  $ &    31.46   &       32   & $    7.56  \pm  0.95 $\\
\hline
\hline\\
\multicolumn{4}{l}{Column 2 gives the average radius of the points 
in the bin.}\\
\multicolumn{4}{l}{Column 3 gives the number of stars per bin.}
\end{tabular}
\end{table}


\end{document}